# Thermal Transport Properties of Bechgaard Salts $(TMTSF)_2PF_6$ and $(TMTSF)_2ClO_4$: Implication of Spin-Charge Separation

Yisheng Chai[*], Hongshun Yang[†], Jian Liu, Chenghai Sun, Huixian Gao, Xudong Chen, Liezhao Cao and Jean-Claude Lasjaunias[1]

Physics Department, University of Science and Technology of China, Hefei, Anhui 230026, P. R. China

[1]Centre de Recherches sur les Tres Basses Temperatures, Laboratoire Associe a l'Universite Joseph Fourier, CNRS, BP 166 38042, Grenoble Cedex 9, France.

We report thermal transport measurements performed on the quasi-one-dimensional Bechgaard salts $(TMTSF)_2ClO_4$ and $(TMTSF)_2PF_6$ along the *a*-direction. For both salts, magnon-drag effects are found to contribute considerably to thermopower above 80 K. These results imply spin-charge separation in the metallic state for both salts. Moreover, a linear temperature-dependent thermal conductivity is found to be unaffected by anion disorder below an anion ordering transition temperature $T_{AO} \sim 25$ K in $(TMTSF)_2ClO_4$.

KEYWORDS: thermal transport properties, Bechgaard salts, spin-charge separation, quasi-one-dimensional, magnon-drag

---

[*] Present address: Department of Physics, Seoul National University
[†] E-mail address: yanghs@ustc.edu.cn

Weakly interacting three-dimensional (3D) electron systems are generally described by the Fermi-liquid theory that charges and spins move together. However, non-Fermi-liquid behaviors in 1D or 2D systems, such as the Luttinger-liquid,[1] exhibit spin-charge separation, that is, collective spin and charge excitations move separately. It has long been sought to verify this novel effect in copper oxide superconductors[2] or 1D quantum wires.[3] Recent progress in angle-resolved photoemission spectroscopy study provided the first direct evidence of spin-charge separation in the quasi-1D compound $SrCuO_2$.[4] Another group of quasi-1D conductors, the Bechgaard salts $(TMTSF)_2X$ (TMTSF= tetramethyltetraselenafulvalene and $X = PF_6$, $ClO_4$, etc.) and their variant TMTTF, where selenium is replaced by sulfur, have been the source of speculation for spin-charge separation. This possibility in Bechgaard salts is based on the fact that, from highly insulating TMTTF to metallic TMTSF salts, there is an abrupt change in their charge-transport properties while their magnetic properties remain nearly unchanged.[5]

Thus, it is expected to observe spin-charge-separation in thermal transport properties as well. In the thermopower of a 1D Luttinger-liquid, the charge diffusion thermopower ($S_d$) has a linear temperature dependence[6,7] while a pure spin contribution has an opposite sign of thermopower at very high temperatures.[8] In addition, the magnon-drag thermopower proposed by Bailyn[9] should be observed owing to the interaction between charge and spin excitations in the Luttinger-liquid. Therefore, TMTSF salts are ideal for this poorly explored subject, although recent thermal conductivity measurements have failed to confirm the presence of spin-charge separation in this system.[10] The few thermopower experiments for TMTSF at high temperatures[11-14] exhibit substantial differences and are not satisfactorily explained by the Fermi-liquid picture.

For TMTSF salts, stacks of TMTSF pile up along the *a*-axis with transfer integrals of $t_a:t_b:t_c$ = 250:25:1 meV. A carrier concentration of one hole/unit cell is confirmed at high temperature by the measurements of Hall coefficient,[15] leading to a three-quarter or half-filled band. However, at a level as low as 10 K, optical conductivity revealed a zero-energy mode with only 1% of the total spectral weight responsible for the large conductivity,[16] together with a gap of approximately 200 cm$^{-1}$. These optical data, combined with the results of spin susceptibility, which are interpreted in terms of the Hubbard model with an exchange constant $J_{ex}$ ≈ 1400 K,[17] are suggestive of spin-charge separation and non-Fermi-liquid behavior in the metallic state of TMTSF salts. Recent reports on magneto-thermoelectric effects in TMTSF salts have revealed giant Nernst signals also near 10 K, casting serious doubts on the Fermi-liquid picture near this temperature (T) region.[18-20] On the other hand, with decreasing *T*, a 1D → 2D dimensional crossover is expected at approximately 100 K owing to considerable interchain coupling,[21] which should result in the appearance of Fermi-liquid behavior below 100 K. In order to resolve such apparent contradiction, a near half-filled narrow band induced by anion-ordering split at a slow cooling rate is proposed in the $X$ = ClO$_4$ salt in order to explain the large Nernst effect and negative thermopower ($S$) below the anion ordering (AO) transition temperature $T_{AO}$ ~ 24 K.[20] It is well known that ClO$_4$ anions are ordered by a slow cooling process through $T_{AO}$ and dimerized along the $b$-direction, yielding a relaxed state with a Fermi surface separated by an AO gap which enters the superconducting ground state at 1.2 K. Rapid quenching could result in a complete anion disordering and subsequently to a spin-density-wave (SDW) ground state below 6 K in this compound. The $X$ = PF$_6$ salt also undergoes a metal-insulator transition, which suggests the presence of the SDW ground state at $T_{SDW} \approx 12$ K. Our previous studies of the $S$ of $X$ = PF$_6$ and ClO$_4$ salts at low temperatures, which revealed a negative $S$ in both compounds and different ground states, support the narrow band picture in this $T$ region without the help of the AO transition.[22,23] Therefore, it would also be meaningful to study the thermal conductivity $\kappa$ under different ground states in $X$ = ClO$_4$ salts in this $T$ region.

In this letter, we report the thermal transport studies of (TMTSF)$_2$PF$_6$ and (TMTSF)$_2$ClO$_4$ from room $T$ down to 6 K and a cooling speed effect study on $\kappa$ near $T_{AO}$ in (TMTSF)$_2$ClO$_4$ along the $a$-axis. The existence of a magnon-drag contribution is revealed in the thermopower above 80 K for (TMTSF)$_2$PF$_6$ and above 45 K for (TMTSF)$_2$ClO$_4$, implying spin-charge separation and thus non-Fermi liquid behavior in these $T$ regions. In the different states of (TMTSF)$_2$ClO$_4$, a near-linear $T$-dependent $\kappa$, which is unaffected by different cooling speeds below $T_{AO}$, is observed.

The schematic in the inset of Fig. 1 shows the experimental setup and follows the steady-state thermopower technique.[24] The $T$ gradient used is 0.1–1 K. A more detailed description can be found in our previous publication.[23] The method for thermal conductivity measurement is similar to that of Torizuka et al.[25] Two crystals are mounted together to increase the magnitude of the signal. For (TMTSF)$_2$ClO$_4$, different cooling speeds were used in cooling the samples from 40 K down to 6 K, and then the measurements were performed while warming the sample.

Figure 1 shows the typical temperature dependence of thermopower $S_a(T)$ in (TMTSF)$_2$ClO$_4$ and (TMTSF)$_2$PF$_6$ along the $a$-axis from 290 K down to 6 K. Above 140 and 100 K for (TMTSF)$_2$ClO$_4$ and (TMTSF)$_2$PF$_6$, respectively, $S_a(T)$ is positive, which indicates hole carriers, and decreases nearly linearly with very large positive intercepts for both salts. As $T$ is lowered, $S_a(T)$ decreases faster and changes signs at 18.5 and 25 K for (TMTSF)$_2$ClO$_4$ and (TMTSF)$_2$PF$_6$, respectively. Our results show substantial differences in the $T$-dependent profiles from those of Bechgaard *et al.*[11] and Mortensen,[13] but are close to those of Choi *et al.*[12]

It can be easily seen that the *T*-dependent $S_a(T)$ of the two salts, before crossing zero, cannot be explained solely by a linear diffusion term. Phonon-drag contribution with a Debye temperature $\theta_D \sim 200$ K usually peaks at $\theta_D/5 \approx 40$ K,[26] where no anomalous feature is found in our data. Thus, phonon-drag contribution is negligible in the *T* region we studied. The spin transport mechanism has to be considered seriously, while a pure spin-contributed thermopower in Luttinger-liquid with a negative sign cannot explain the large positive intercepts in $S_a(T)$ near room temperature.

As a result, magnon-drag contribution is suggested in this system and it should be antiferromagnetic in nature.[17] In analogy to the phonon-drag contribution,[26] the magnon-drag thermopower $S_m$ can be written as

$$S_m = \frac{C_m}{dn_0 e}\left(\frac{1/\tau_{mc}}{1/\tau_{mc} + 1/\tau_{mm}}\right), \tag{1}$$

where $C_m$ is the magnon specific heat, $e$ is the unit of charge, $d$ is the dimension of the spin system, $n_0$ is the carrier density participating in the dc transport, and $\tau_{mc}$ and $\tau_{mm}$ are the magnon relaxation times due to magnon-charge scattering and all other magnon scatterings, respectively. In the spin half Heisenberg chains, $C_m$ can be described by the relation[27]

$$\frac{C_m(T)}{Nk_B} = \frac{2T}{3J_{ex}}\left\{1 + \frac{3}{(2L)^3} + O\left[\frac{1}{(2L)^4}\right]\right\} + \frac{2(3^{5/2})}{5\pi}\left(\frac{T}{J_{ex}}\right)^3\left\{1 + O\left(\frac{1}{2L}\right)\right\}, \tag{2}$$

where $L$ is related to the logarithmic correction term and $N$ is the number of approximate spins in the antiferromagnetic chain. Since the spin structure of $(TMTSF)_2PF_6$ can be explained by a linear Hubbard model with parameters of Coulomb repulsion $t_a/U > 0.2$ and $J_{ex} \approx 1400$ K,[17] eq. (2) can be simplified to $3C_m(T)/2Nk_B = T/J_{ex} + 2.9772(T/J_{ex})^3$ for $T \ll J_{ex}$. Moreover, as reported by Bourbonnais and Jerome,[28] the spin susceptibilities of both crystals exhibit very similar $T$ dependent profiles, implying a similar $J_{ex}$ for the $ClO_4$ salt. At the same time, similarly to that for the phonon-drag effect, a simple power law $T$ dependence as the first approximation for $\tau_{mc}$ and $\tau_{mm}$ is assumed and the relaxation term in eq. (1) will have the form $T_0^\beta/(T_0^\beta + T^\beta)$, where $T_0$ is the characteristic temperature.

According to the above discussion, the total thermopower $S_a(T)$ of TMTSF salts has the form

$$S_a(T) = S_d(T) + S_m(T) = AT + B(T + 2.9772T^3/J_{ex}^2)T_0^\beta/(T_0^\beta + T^\beta), \qquad (3)$$

where $AT$ represents the linear diffusion term $S_d(T)$ and $B = 2Nk_B/3dn_0eJ_{ex}$. An excellent agreement can be seen between the experiment data and fitting curves above 80 K for $(TMTSF)_2PF_6$ and above 45 K for $(TMTSF)_2ClO_4$, respectively (thin solid lines in Fig. 1). The best fitting parameters are listed in Table I. The slopes of $S_d$ (=$A$) are close to 0.07 $\mu$V/K$^2$ for both salts, corresponding to a total bandwidth of about 1 eV in a three-quarter-filled 1D tight-binding band theory,[11] consistent with the theoretical value $4t_a$ and the optical measurements. $J_{ex}$ is found to be 1260±30 and 1380±50 K for $(TMTSF)_2ClO_4$ and $(TMTSF)_2PF_6$, respectively. These values are very close to the expected 1400 K, in good agreement with the susceptibility results.[17] On the other hand, there are considerable differences in $T_0$ and $\beta$ between the two salts, where $T_0$ corresponds to $T$ where $\tau_{mc} = \tau_{mm}$. Between the two salts, in terms of their similar $A$ and $J_{ex}$ values, which implies similar $\tau_{mc}$ values, $\tau_{mm}$ should have different power law behaviors ($\beta$ is different) and give rise to different $T_0$ values. Finally, from $B$ and $J_{ex}$, we can obtain the dimensionless ratio $N/n_0 = 3deJ_{ex}B/2k_B$ where $d = 1$, which links the effective number of spins to that of charges and is calculated to be 3.76 and 1.73 for $(TMTSF)_2ClO_4$ and $(TMTSF)_2PF_6$, respectively. We will later show that these values are consistent with the susceptibility data of the two salts.

Generally, the charge transport properties are comparable in both salts. The main difference in their thermopower comes from the magnitude and $T$-dependent behavior of the magnon-drag contribution, where different magnon scattering mechanisms and probably the ratio of the number of spins and charges more or less play a role. Note that the magnitude of the magnon drag can be suppressed either by impurities or microcracks created during cooling, which will lead to a decrease in $\tau_{mm}$. In this respect, we can understand the data reported in refs. 11-14 within our model, but with a smaller magnitude of magnon-drag contribution. Particularly, the $S_d$ values are very similar to our results.

Before discussing the low-$T$ data, we would like to mention the large difference in the ratio of $N/n_0$ between the two salts. In particular, the charge term $S_d$ of both salts implies a three-quarter-filled band and an $n_0$ of one hole/unit cell. The $N$ of $(TMTSF)_2ClO_4$ should be much larger than that of $(TMTSF)_2PF_6$ accordingly. We may argue that it has already been revealed in the spin susceptibility data. From TMTSF to TMTTF salts, there is a threefold enhancement of spin susceptibility in terms of the increase in localization.[17] Similarly, from $(TMTSF)_2ClO_4$ to $(TMTSF)_2PF_6$, an increase in the magnitude of susceptibility should be observed if they have the same $N$. However, it is in strong contrast to the larger magnitude of susceptibility in $(TMTSF)_2ClO_4$.[28] Such a contradiction could only be explained by a much larger $N$ in $(TMTSF)_2ClO_4$. Nevertheless, such an increase in $N$ in $(TMTSF)_2ClO_4$ found in both $S_a$ and susceptibility must have the same physical origin.

We now turn to the lower $T$ region where $S_a(T)$ deviates significantly from the fitting curves and shows sign changes, as shown in Fig. 2. The $S_a(T)$ of $(TMTSF)_2PF_6$, above $T_{SDW} = 12.5$ K, presents a shallow valley at 13.7 K, in accordance with previous data.[12] Negative $S_a$ values for both compounds and different ground states were observed, as shown in Figs. 2 and 3(a), respectively. A more detailed discussion on the data of $(TMTSF)_2PF_6$ and $(TMTSF)_2ClO_4$ below 30 K is given in ref. 23 and ref. 22, respectively. We showed more cooling speed data in Fig. 3(a) for clarity. It can be easily seen that the negative $S_a$ in $(TMTSF)_2PF_6$ which has no AO transition, also supports the picture of the effective doping of interchain couplings to the correlated zero-energy modes found in optical conductivity in ref. 16.

To understand the paradox of the Fermi-liquid near this $T$ region, the $T$-dependent thermal conductivity $\kappa_a(T)$ along the $a$-direction for $(TMTSF)_2ClO_4$ with different cooling speeds were measured from 40 to 18 K, as shown in Fig. 3(b). The data, since the spin and charge contributions to the thermal conductivity $\kappa_a$ under different cooling speeds in $(TMTSF)_2ClO_4$ may react differently, are expected to reveal if $(TMTSF)_2ClO_4$ is indeed a non-Fermi-liquid at this $T$-region. The absolute value of $\kappa_a$ varies by about ±50% owing to the inaccuracy of the sample geometry measurements; fortunately, as will be seen later, this does not affect our discussion below. Our findings are rather surprising in that different cooling speeds have no discernable effect on $\kappa_a$ compared with the significant influence on $S_a$. In general, heat conduction has three possible origins: electrons, phonons and magnons. As in the case of the $X = ClO_4$ salt, first, the pure electronic contribution $\kappa_{ch}$ estimated by the Wiedemann-Franz law $\kappa_{ch}(T) = L_0 T \sigma(T)$ can only account for a small fraction of total thermal conductivity within this $T$ region. In particular, even though a speed of 0.075K/s will not make a sample entering a completely quenched state, it is enough to demonstrate a clear deviation from the relaxed state in resistivity, and subsequently a difference in $\kappa_{ch}$ below $T_{AO}$. Second, low-$T$ data is close to the linear behavior below about 30 K, as shown in Fig. 3(b). Actually, in our previous specific heat measurement in (TMTSF)$_2$ClO$_4$ in ref. 29, we found that its specific heat consists of an acoustic $T^{2.7}$ contribution and two Einstein modes below 15 K. However, if the thermal conductivity is dominated by phonon conductivity $\kappa_{ph}$, similarly to the analysis and data used in ref. 30 for estimating the phonon mean free path $l_{ph}$ at 25 K with $\kappa_{ph} = 1/3\ C_{ph}v_s l_{ph}$: where $C_{ph}$ is lattice specific heat, $C_{ph}/T^3\ (T = 25\ \text{K}) \approx 6$ mJ/(mol·K$^4$),[29] $v_s$ = 3000 m/s (velocity of sound) and $\kappa_{ph}(25\ \text{K}) \approx 3$ W/(K·m) in our data. $l_{ph}$ would be only 0.0148 $\mu$m at 25 K. Third, likewise, the 1D magnon mean free path $l_m$ can be estimated similarly using $\kappa_m = C_m v_m l_m$,[10] where $C_m$ is nearly linear in this $T$ region, $v_m = (\pi^2 k_B/h)J_{ex}a \approx 1.75 \times 10^5$ m/s for the velocity of magnons ($k_B$ is the Boltzmann constant, $h$ is the Planck constant and $a \approx 0.7$ nm is the lattice constant along the chains). Let us assume that the $T$ linear dependence of $\kappa_a$ indicates a nearly invariable $l_m$ near $T_{AO}$, which is consistent with the expectation that the magnon is much less affected by an anion disorder in (TMTSF)$_2$ClO$_4$ than the charge transport. We can obtain $l_m$ to be 64 $\mu$m within the $T$ region we discussed with a nice fit between $\kappa_m$ and experimental data below 30 K, as shown in Fig. 3(b). Meanwhile, we find that $C_m/T^3$ is only about 0.016 mJ/(mol·K$^4$) at 25 K. Therefore, it seems that, even though a phonon makes a predominant contribution in heat capacity its contribution to $\kappa_a$ is small which requires an $l_{ph}$ smaller than 0.0148 $\mu$m at 25 K in our sample. However, after considering the linear $T$-dependent behavior of $\kappa_a$ and a considerable cooling-speed-dependent lattice specific heat found in ref. 22, we could still rule out the phonon-dominated $\kappa_a$. These facts taken together imply the existence of spin-charge separation in this compound.

In conclusion, we measured the $T$-dependent $S_a$ in Bechgaard salts (TMTSF)$_2$ClO$_4$ and (TMTSF)$_2$PF$_6$ along the $a$-direction. In addition, the role of the cooling speed effect is compared between $S_a$ and $\kappa_a$ near $T_{AO}$ in (TMTSF)$_2$ClO$_4$. These results are compatible with the presence of spin-charge separation at high $T$ in both salts and in (TMTSF)$_2$ClO$_4$ near $T_{AO}$.

**Acknowledgements**: We thank N. H. Andersen and G. R. Stewart for review of the manuscript and helpful discussion. This work is supported by the National Natural Science Foundation of China (Grant No. 10374082)

**Figure Captions:**

Fig. 1. Temperature-dependent thermopower $S_a(T)$ of $(TMTSF)_2ClO_4$ (triangle) and $(TMTSF)_2PF_6$ (circle) single crystals. The solid lines are the fitting curves from eq. (3) for both crystals. The dotted and dashed lines are the fitting results of the diffusion thermopower $S_d$ (for details see the text). The inset shows the schematic diagram of experimental setup. *C1* and *C2* are large fixed Cu plates controlled at different temperatures. *C3* is a small Cu sheet residing on two loosely suspended wires so that it could move freely along the *a*-direction and is thermally connected with *C1* with 20 gold wires.

Fig. 2. Temperature-dependent thermopower $S_a(T)$ below 40 K for both salts.

Fig. 3. Temperature-dependent thermal transport measurements of $(TMTSF)_2ClO_4$ under different cooling speeds near the anion-ordering temperature $T_{AO}\sim25$ K. (a) Thermopower $S_a$ values at cooling speeds of 0.0005 K/s (square), 0.005 K/s (downward triangle), 0.022 K/s (circle), 0.3 K/s (upward triangle) and 1 K/s (solid line), respectively. (b) Thermal conductivities $\kappa_a$ at cooling speeds of 0.0005 K/s (circle) and 0.075 K/s (triangle), respectively. The solid line is a fit of $\kappa_m$ to the data below 30 K.

Table I. List of fitting parameters for (TMTSF)$_2$ClO$_4$ and (TMTSF)$_2$PF$_6$ from eq. (2): *A*, *B*, exchange interaction $J_{ex}$, $T_0$, $\beta$ and $(N/n_0)/d$, respectively, where $(N/n_0)/d = 3eJ_{ex}B/2k_B$.

|  | $A$ ($\mu$V/K$^2$) | $B$ ($\mu$V/K$^2$) | $J_{ex}$ (K) | $T_0$ (K) | $\beta$ | $(N/n_0)/d$ |
|---|---|---|---|---|---|---|
| ClO$_4$ | 0.079±0.001 | 0.17±0.01 | 1260±30 | 114.4 | 2.5±0.1 | 3.76 |
| PF$_6$ | 0.076±0.001 | 0.07±0.01 | 1380±50 | 182.3 | 3.8±0.4 | 1.73 |

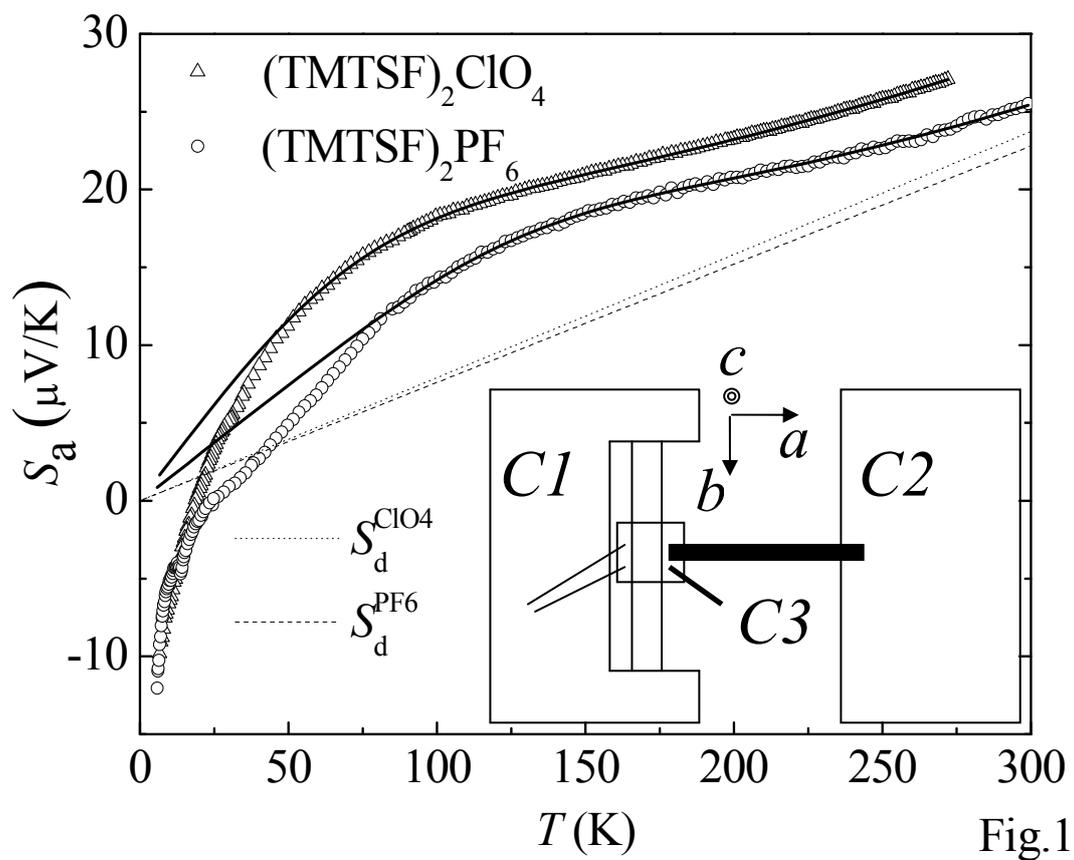

Fig.1

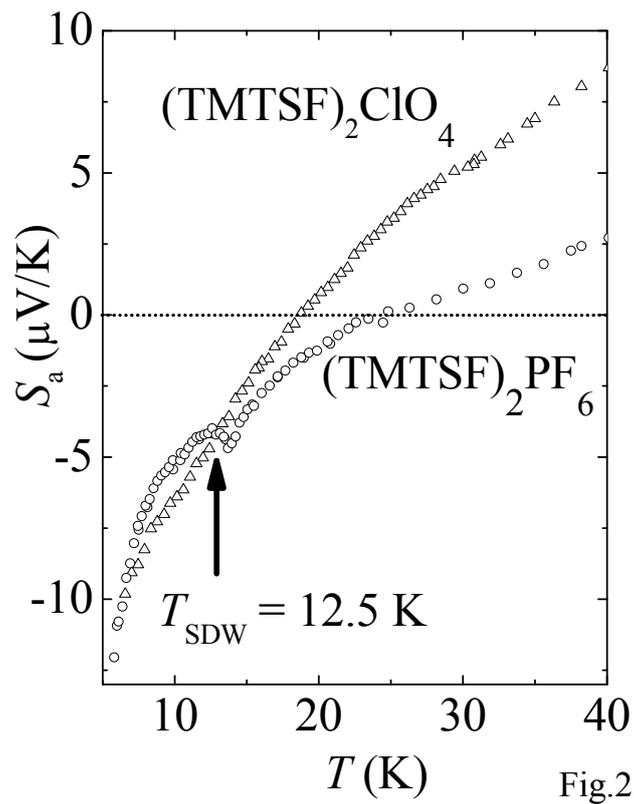

Fig.2

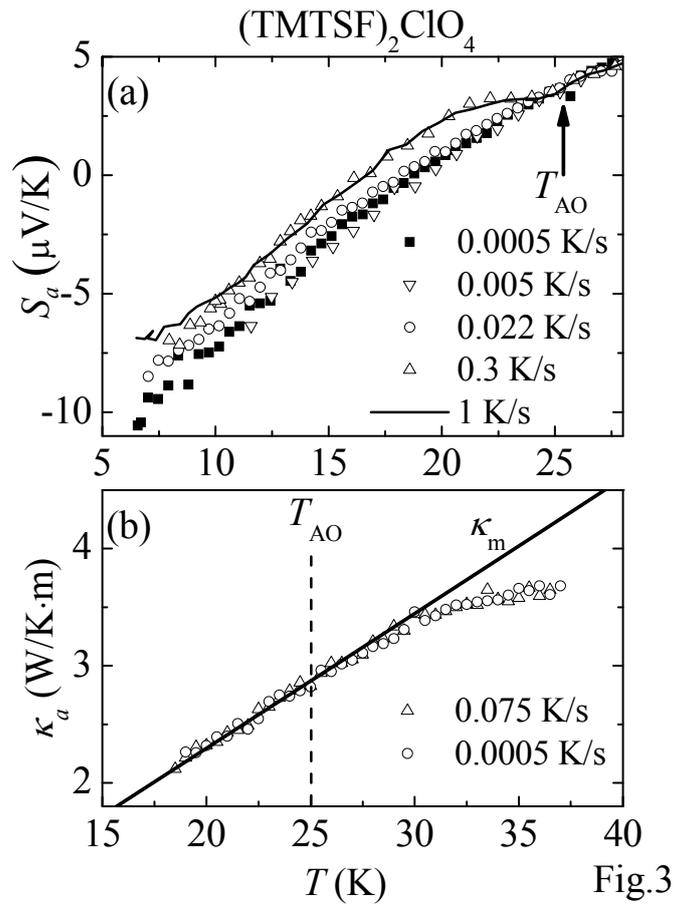

Fig.3